\documentclass[5p,a4paper]{elsarticle}
\biboptions{sort&compress,numbers}
\usepackage[bookmarks=true,colorlinks=true,urlcolor=blue,linkcolor=blue,citecolor=blue]{hyperref} 
\usepackage{units}
\usepackage{txfonts}
\usepackage{multirow}
\journal{Computational and Theoretical Chemistry}

\usepackage[labelfont=bf]{caption}
\begin{document}

\begin{frontmatter}

\title{Coverage Dependence of the Level Alignment for Methanol on TiO$_{\mathrm{2}}$(110)}

\author[ICN2]{Annapaola Migani\corref{corresponding}}
\address[ICN2]{ICN2 - Institut Catal\`{a} de Nanoci\`{e}ncia i Nanotecnologia and CSIC - Consejo Superior de Investigaciones Cientificas, ICN2 Building, Campus UAB, E-08193 Bellaterra (Barcelona), Spain}
\ead{annapaola.migani@cin2.es}

\author[Nano-Bio]{Duncan J. Mowbray\corref{corresponding}}

\address[Nano-Bio]{Nano-Bio Spectroscopy Group and ETSF Scientific Development Center, Departamento de F{\'{\i}}sica de Materiales, Universidad del Pa{\'{\i}}s Vasco UPV/EHU and DIPC, E-20018 San Sebasti\'{a}n, Spain}
\ead{duncan.mowbray@gmail.com}

\cortext[corresponding]{Corresponding authors. Address: ICN2 Building, Campus UAB, E-08193 Bellaterra, Tel.: +34 937 37 3605 (Annapaola Migani). Address: Centro Joxe Mari Korta,  Avenida de Tolosa 72, E-20018 San Sebasti\'{a}n, Tel.: +34 943 01 8392 (Duncan J. Mowbray).}


\begin{abstract}
\hspace{-0.2cm}\begin{tabular*}{20cm}{p{5cm}p{12.9cm}}
\noindent\textsc{\large{a r t i c l e \hspace{0.1cm}  i n f o}}

\normalsize
\noindent\rule{4.5cm}{0.4pt} 
&
\noindent\textsc{\large{a b s t r a c t}}

\noindent\par\rule{12.9cm}{0.4pt}\\
\footnotesize

\textit{Article history:}

Submitted \today

\noindent\rule{4.5cm}{0.5pt} 

HOMO

Wet electron level

Two photon photoemission

Ultraviolet photoemission spectroscopy

Photocatalysis
&
\small
Electronic level alignment at the interface between an adsorbed molecular layer and a semiconducting substrate determines the activity and efficiency of many photocatalytic materials.  We perform $G_0W_0$ calculations to determine the coverage dependence of the level alignment for a prototypical photocatalytic interface: \nicefrac{1}{2} and 1 monolayer (ML) intact and dissociated CH$_3$OH on rutile TiO$_2$(110).   We find changes in the wavefunction's spatial distribution, and a consequent renormalization of the quasiparticle energy levels, as a function of CH$_3$OH coverage and dissociation.  Our results suggest that the occupied molecular levels responsible for hole trapping are not those observed in the ultraviolet photoemission spectroscopy (UPS) spectrum. Rather, they are those of isolated CH$_3$O on the surface. We find the unoccupied molecular levels have either 2D character with weight above the surface at 1 ML coverage, or significant hybridization with the surface at \nicefrac{1}{2} ML coverage. These results suggest the resonance observed in the two photon phooemission (2PP) spectrum arises from excitations to unoccupied ``Wet electron'' levels with 2D character.
\end{tabular*}
\end{abstract}

\end{frontmatter}


\section{Introduction}

TiO$_2$ is one of the most technologically important photocatalytic materials \cite{YatesChemRev,HendersonSurfSciRep}. Since the first report in the early 1970s of photocatalytic H$_2$O splitting on TiO$_2$ electrodes \cite{FujishimaNature}, this material has been extensively investigated experimentally \cite{FujishimaReview}.  This is primarily due to its application in molecular hydrogen (H$_2$) energy technology \cite{Crabtree,DOE}. In 1980,
photocatalyzed H$_2$ generation was shown to be remarkably enhanced when
sacrificial CH$_3$OH, a hole trap, was added to H$_2$O \cite{Kawai1980}. 

Since then, the
photocatalytic chemistry of CH$_3$OH on TiO$_2$ has attracted an increasing
amount of attention. In particular, model surface science experiments have been conducted on the single crystal rutile TiO$_2$(110) surface, under ultrahigh vacuum (UHV) and controlled coverage conditions.  Experimentalists have characterized the electronic structure using ultraviolet, X-ray, and two photon photoemission spectroscopy (UPS, XPS, and 2PP) \cite{Onishi198833,UPSMethanol1998,Weixin,Onda200532,PetekScienceMethanol,PetekScienceH2O,MethanolSplitting2010,Methanol2PPYang}, scanning tunnelling microscopy (STM) \cite{MethanolSplitting2010,Methanol2PPYang,HendersonReview,Henderson2012}, and reaction products using temperature programmed desorption (TPD) \cite{Weixin,Henderson2011,MethanolPhotocatalysis,Friend2012,MethanolSplitting2013,HydrogenfromMethanolPhotocatalysis}.  These experiments have shown that CH$_3$OH can be photocatalytically dissociated on TiO$_2$(110) to form CH$_3$O and surface OH species \cite{MethanolSplitting2010,MethanolSplitting2013}, and selectively photooxidized to CH$_2$O \cite{MethanolPhotocatalysis} and HCOOCH$_3$ \cite{Weixin,Friend2012}.  Moreover, the formation of CH$_2$O can be accompanied by the photocatalized production of H$_2$  \cite{HydrogenfromMethanolPhotocatalysis}. A similar photocatalytic dissociation of H$_2$O on
TiO$_2$(110) has been recently reported \cite{WaterDissociationJACS2012Jin}, although with a much lower efficiency.

The mechanism for these reactions is still unclear and requires
an accurate theoretical treatment. The reaction begins upon irradiating
the TiO$_2$ photocatalyst with UV light at energies exceeding the band gap.  This leads to the formation of electron-hole pairs. To understand the
reaction mechanism, it is necessary to study the
interaction of the photogenerated electron-hole
pairs with adsorbed CH$_3$OH and H$_2$O.
In particular, the energy level alignment of the occupied O 2p surface and frontier molecular levels determines both the mechanism and activity of the photoxidation processes (e.g., CH$_2$O or HCOOCH$_3$ formation). On the other hand, the level alignment of the unoccupied Ti 3d surface and frontier molecular levels determines the mechanism and activity of the photoreduction process (e.g., H$_2$ formation). 

The level alignment of CH$_3$OH on TiO$_2$(110) has been probed by UPS for the occupied levels \cite{Onishi198833}, and 2PP for the unoccupied levels \cite{Onda200532}.  Based on quasiparticle (QP) $G_0W_0$ calculations for one monolayer (ML) of CH$_3$OH on TiO$_2$(110), we recently showed that the molecular structures measured in both UPS \cite{Onishi198833} and 2PP \cite{Onda200532} experiments are predominantly intact CH$_3$OH \cite{OurJACS}.   For 1 ML \nicefrac{1}{2}-dissociated CH$_3$OH layers, we found the highest occupied molecular orbitals (HOMO) of CH$_3$OH are nearer the valence band maximum (VBM). This more favorable HOMO alignment may explain  why CH$_3$O is more photocatalytically active than CH$_3$OH \cite{OurJACS}.  

\begin{figure}[!t]
\includegraphics[width=\columnwidth]{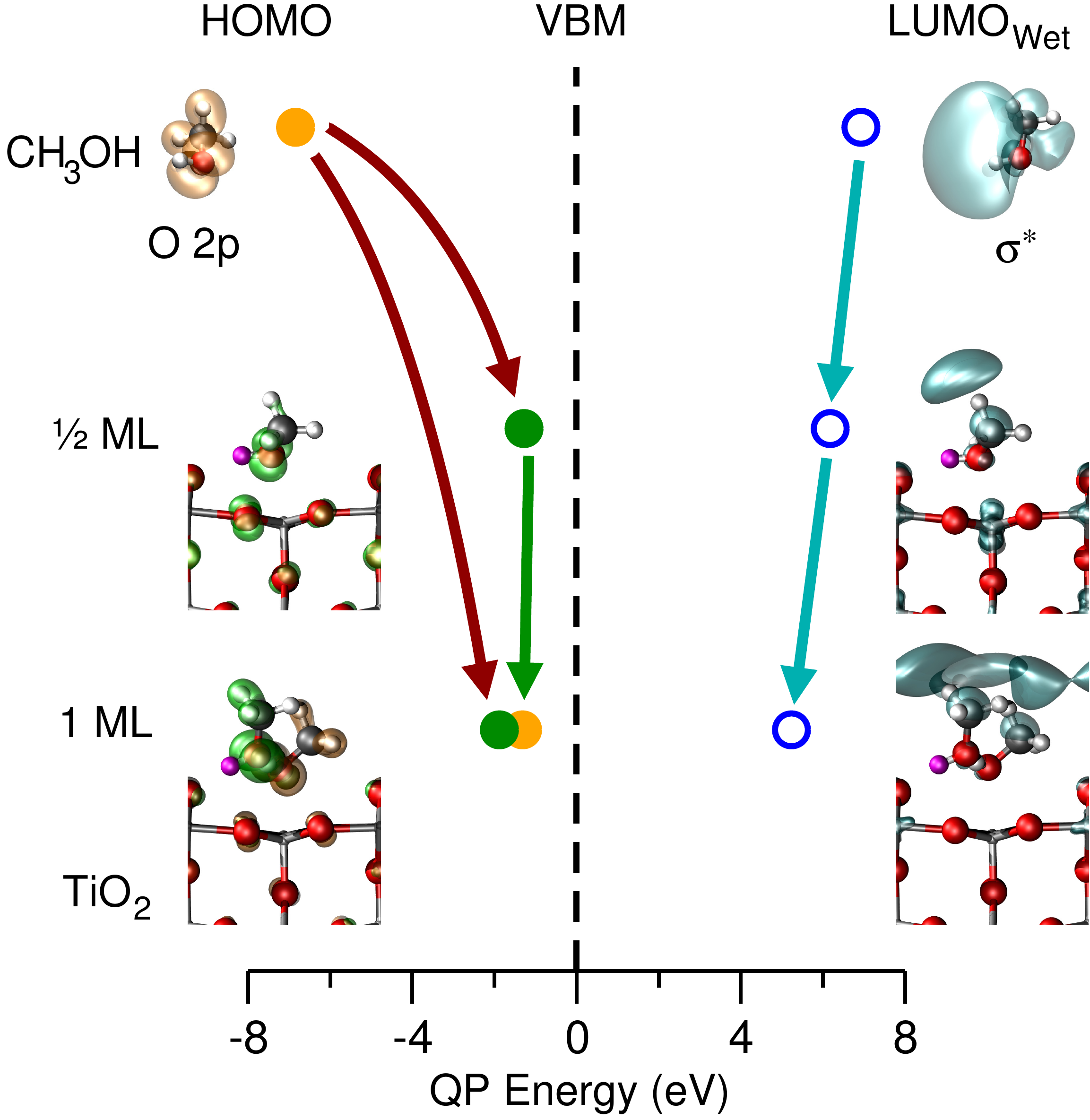}
\caption{Renormalization of the QP energies of HOMO and LUMO$_{\mathrm{Wet}}$   levels in the gas phase due to adsorption on TiO$_2(110)$ for intact CH$_3$OH \nicefrac{1}{2} ML and 1 ML \cite{OurJACS} coverages.  H, C, O, Ti, and H$^+$ atoms are represented by white, grey, red, silver, and magenta balls, respectively. 
}
\label{fgr:schematic}
\end{figure}

However, these experiments may incorporate lower coverage domains.  Here, we will probe how the molecular coverage can directly modify the type, spatial distribution, and QP energy of the HOMO for intact and dissociated CH$_3$OH on TiO$_2$(110).  These are predominantly non-bonding O 2p orbitals, with some C--H $\sigma$ character. The HOMO of the dissociating CH$_3$OH (HOMO$_{\mathrm{D}}$) and the non-dissociating CH$_3$OH (HOMO$_{\mathrm{ND}}$) are shown in Fig.~\ref{fgr:schematic} for CH$_3$OH in gas phase and \nicefrac{1}{2} ML and 1 ML intact CH$_3$OH on TiO$_2$(110).

In 2PP experiments of CH$_3$OH on TiO$_2$(110), an excited resonance peak has been independently observed at about $2.4$~eV \cite{Onda200532,PetekScienceMethanol,MethanolSplitting2010,Methanol2PPYang} above the Fermi level $\varepsilon_F$. 
This resonance is due to excitations from reduced Ti$^{+3}$ 3d levels located $\sim 0.7$~eV below the Fermi level through an intermediate level to the vacuum using a 3.05~eV light source \cite{PetekScienceMethanol}.
However, which intermediate electronic level is responsible for the observed 2PP resonance is still under debate.

\begin{figure}[!t]
\includegraphics[width=\columnwidth]{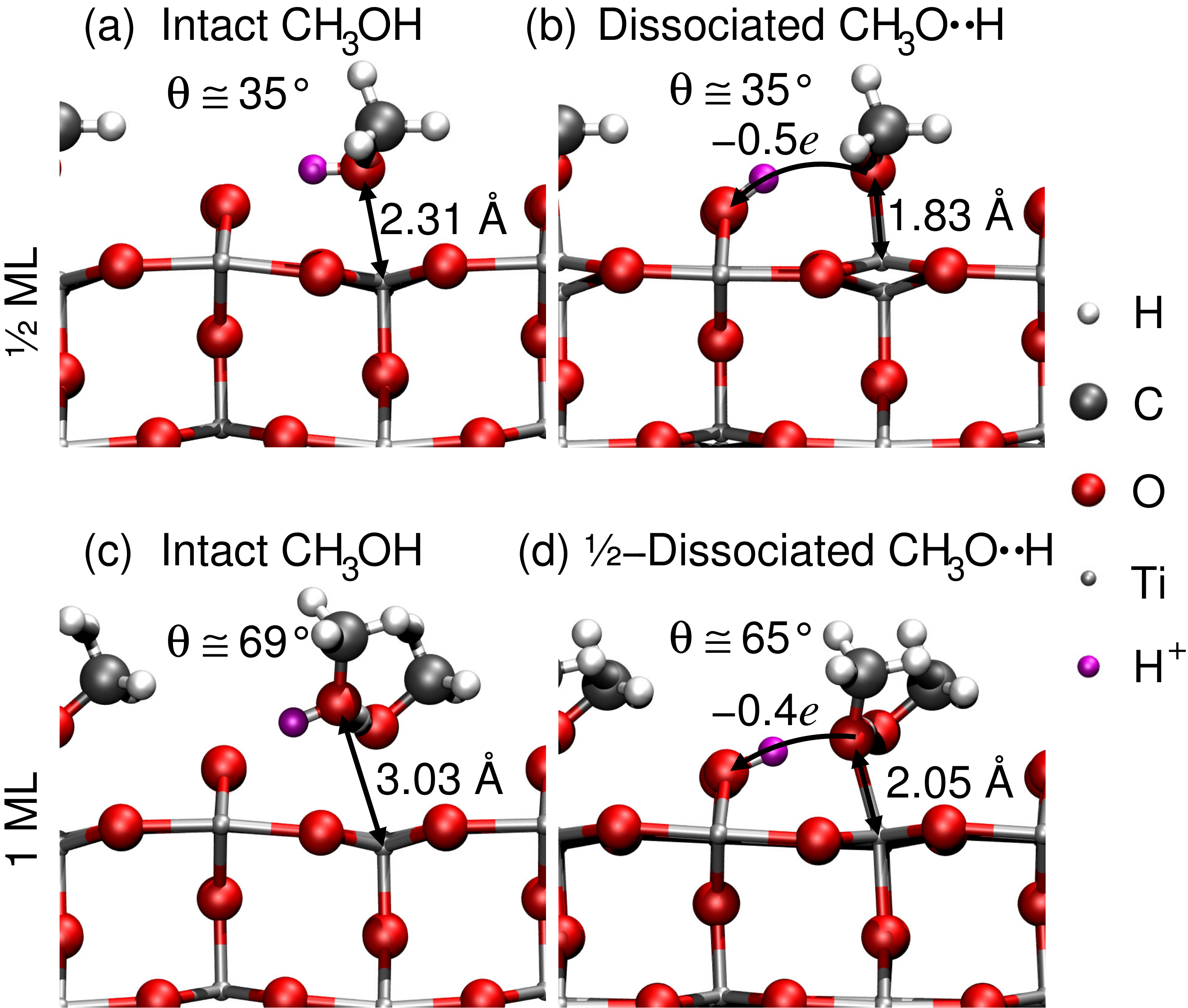}
\caption{The atomic structures of (a,b) \nicefrac{1}{2} ML and (c,d) 1 ML \cite{Jin,OurJACS} of (a,c) intact, (b) dissociated, and (d) \nicefrac{1}{2}-dissociated CH$_3$OH on TiO$_2$(110). The transferred proton is marked in magenta, along with the accompanying charge transfer of (b) 0.5 and (d) 0.4 electrons.  Relevant geometrical parameters of the dissociating CH$_3$OH, i.e. O--Ti distance in \AA\ and molecular tilt $\theta$ in $^{\circ}$ relative to the surface, are also provided. }
\label{fgr:structures}
\end{figure}

On the one hand, H.\ Petek and coworkers \cite{PetekScienceMethanol} attribute the 2PP resonance to excitations through a ``Wet electron'' level.  This electronic level has a two dimensional (2D) $\sigma*$ character associated with the CH$_3$OH C--H bonds, and significant weight above the H atoms outside the molecular layer \cite{PetekScienceMethanol,OurJACS,MiganiLong,Jin}.  They report this resonance attains its maximum intensity at 1 ML coverage \cite{Onda200532}.  The lowest unoccupied molecular orbitals with Wet electron character (LUMO$_{\mathrm{Wet}}$) are shown in Fig.~\ref{fgr:schematic} for CH$_3$OH in gas phase and intact \nicefrac{1}{2} ML and 1 ML coverages on TiO$_2$(110).

On the other hand, X.\ Yang and coworkers \cite{MethanolSplitting2010,Methanol2PPYang} attribute the 2PP resonance to the presence of dissociated CH$_3$OH, i.e., CH$_3$O, on TiO$_2$(110). Specifically, they state the 2PP resonance is from 3d levels of CH$_3$O occupied cus Ti sites. They find the 2PP resonance significantly increases in intensity and shifts to lower energy with illumination \cite{MethanolSplitting2010,Methanol2PPYang}. For this reason, the 2PP resonance appears to be a consequence of CH$_3$OH photodissociation upon UV irradiation \cite{MethanolPhotocatalysis,MethanolSplitting2010}.  Moreover, they report a similar resonance is observed at lower coverages (\nicefrac{1}{2} ML and \nicefrac{1}{6} ML).  

At such low coverages, the $\sigma^*$ levels of neighboring CH$_3$OH are not expected to interact.  This means the Wet electron levels lose their 2D character, and should be destabilized. It is thus unclear whether a Wet electron level would be accessible at the 2PP resonance energy (2.4~eV) and lower coverages (\nicefrac{1}{2} ML and \nicefrac{1}{6} ML) \cite{MethanolPhotocatalysis,MethanolSplitting2010}.  For this reason, X.\ Yang claims the 2PP resonance cannot be interpreted as a Wet electron level, as the latter is expected to be accessible only at higher coverages (1 ML) \cite{MethanolSplitting2010,Methanol2PPYang}. 

Within the context of UPS and 2PP experiments, it is important to understand how the energy and spatial distribution of the CH$_3$OH HOMO and Wet electron levels change with the surface coverage and following dissociation.  However, an accurate description of the interfacial level alignment requires a QP $G_0W_0$ treatment.  This is because the screening of the electron--electron interaction in photocatalytic interfacial systems has a strong spatial depedence \cite{OurJACS,MiganiLong,GiustinoPRL2012, NaClCorrelation, RenormalizationLouie, JuanmaRenormalization1}.  In other words, interfacial levels are screened according to their weight in the bulk, molecular, and vacuum regions \cite{OurJACS,MiganiLong}.

In this paper, we perform $G_0W_0$ calculations for the \nicefrac{1}{2} ML intact and dissociated and the 1 ML intact and \nicefrac{1}{2}-dissociated CH$_3$OH on TiO$_2$(110) structures shown in Fig.~\ref{fgr:structures}.  1 ML coverage corresponds to the adsorption of one CH$_3$OH per coordinately unsaturated (cus) Ti site, while \nicefrac{1}{2} ML coverage corresponds to one CH$_3$OH for every other cus Ti site of the TiO$_2$(110) surface.  In so doing, we show how the HOMO and LUMO$_{\mathrm{Wet}}$ CH$_3$OH QP energies are renormalized upon absorption and increasing coverage on TiO$_2$(110) (\emph{cf.} Fig.~\ref{fgr:schematic}).

\section{Computational methods}

The QP $G_0W_0$ approach involves the single-shot correction of the Kohn-Sham (KS) density functional theory (DFT) eigenvalues by the self energy $\Sigma = i G W$, where $G$ is the Green's function and $W$ is the screening \cite{GW}. $W$ is obtained from  the  dielectric function, based on the KS wavefunctions \cite{AngelGWReview}.  This is calculated using linear response time-dependent DFT within the random phase approximation (RPA), including local field effects \cite{KresseG0W0}.  Through the $G_0W_0$ technique one obtains a first-order approximation to the QP eigenvalues.  However, the vacuum level $E_{\mathit{vac}}$ and wavefunctions used are those from standard DFT. 

All calculations have been performed using the DFT code \textsc{vasp}  within the projector augmented wave (PAW) scheme \cite{kresse1999}. We used a generalized gradient approximation (PBE) \cite{PBE} for the xc-functional \cite{kresse1996b}.  The geometries have been fully relaxed, with all forces $\lesssim$ 0.02 eV/\AA, a plane-wave energy cutoff of 445 eV, an electronic temperature $k_B T\approx0.2$ eV with all energies extrapolated to $T\rightarrow 0$ K, and a PAW pseudopotential for Ti which includes the 3$s^2$ and 3$p^6$ semi-core levels \cite{HybertsenTiO2GW,AmilcareTiO2GW}.

\begin{figure*}[!t]
\includegraphics[width=\textwidth]{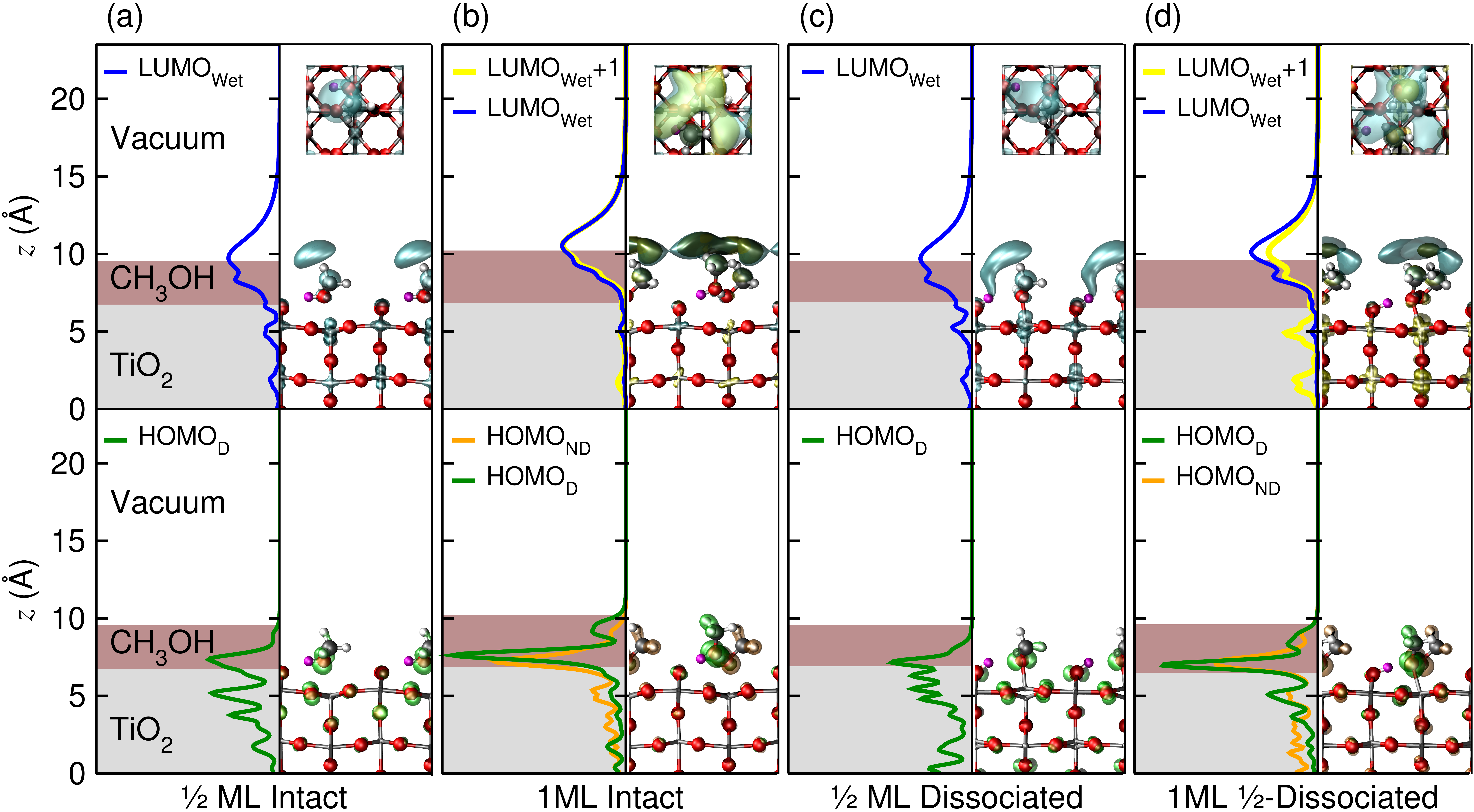}
\caption{Calculated  LUMO$_{\mathrm{Wet}}$ (blue), LUMO$_{\mathrm{Wet}}+1$ (yellow), HOMO$_{\mathrm{D}}$ for the dissociating CH$_3$OH (green), and HOMO$_{\mathrm{ND}}$ for the non-dissociating CH$_3$OH (orange) isosurfaces and their average densities in the $x y$-plane at $\Gamma$ versus distance $z$ in \AA\ from the center of the TiO$_{2}$ substrate for (a,b) intact, (c) dissociated,  and (d) \nicefrac{1}{2}-dissociated (a,c) \nicefrac{1}{2} ML and (b,d) 1 ML \cite{OurJACS} CH$_3$OH on TiO$_{{2}}$(110).  TiO$_2$ bulk, CH$_{{3}}$OH molecular layer, and vacuum are depicted by grey, brown, and white regions, respectively.
H, C, O, Ti, and H$^+$ atoms are represented by white, grey, red, silver, and magenta balls, respectively. 
}
\label{fgr:wavefunctions}
\end{figure*}

We have modelled the atomic structures of \nicefrac{1}{2} ML intact CH$_3$OH and its dissociated counterpart on TiO$_2$(110), as shown in Fig.~\ref{fgr:structures}(a) and (b).  For the 1 ML coverage we have used the most stable structure of intact CH$_3$OH and its \nicefrac{1}{2}-dissociated counterpart on TiO$_2$(110) \cite{Jin} shown in Fig.~\ref{fgr:structures}(c) and (d).
In each case we have included adsorbates on both sides of a four layer slab, with $C_{2}$ symmetry.  We used a $1\times2$ unit cell of $6.497 \times 5.916 \times 47.0$~\AA$^3$, corresponding to the experimental lattice parameters for bulk rutile TiO$_{\mathrm{2}}$ \cite{TiO2LatticeParameters} in the surface plane.  This provides $\gtrsim 27$~\AA\ of vacuum between repeated images.

We employed a $\Gamma$ centered $4\times4\times1$ \textbf{k}-point mesh, with 760 bands for \nicefrac{1}{2} ML and 880 bands for 1 ML coverages, i.e.\ $\sim$9.2 unoccupied bands per atom, an energy cutoff of 80 eV for the number of \textbf{G}-vectors, and a sampling of 80 frequency points for the dielectric function.  We have previously shown that these parameters provide an accurate description of the level alignment for 1 ML of CH$_3$OH on TiO$_2$(110) \cite{OurJACS} and the bulk rutile TiO$_2$ electronic band gap and optical spectrum \cite{MiganiLong}.  

The alignment of the experimental UPS and 2PP spectra for CH$_3$OH on TiO$_2$(110) has been performed relative to the VBM. This is accomplished by shifting by $+3.2$~eV the UPS and 2PP spectra, which are measured relative to the Fermi level. A more detailed discussion of the experimental energy references is provided in Ref.~\citenum{MiganiLong}.

\section{Results and discussion}

We have performed $G_0W_0$ QP calculations for the structures shown in Fig.~\ref{fgr:structures}. These correspond to \nicefrac{1}{2} ML (intact and dissociated) and 1 ML (intact and \nicefrac{1}{2}-dissociated) of CH$_3$OH on TiO$_2$(110). 

For the \nicefrac{1}{2} ML structures (a,b) CH$_3$OH is adsorbed at the cus Ti site of the  TiO$_2$(110) surface. For the intact structure, CH$_3$OH forms an interfacial hydrogen bond with a neighboring bridging O atom (O$_{\mathit{br}}$) of the surface.  For the dissociated structure, CH$_3$OH transfers a proton and an accompanying charge of 0.5 electrons to the O$_{\mathit{br}}$. 

The 1 ML structures (c,d) have an additional CH$_3$OH at an adjacent cus Ti site. This non-dissociating CH$_3$OH forms an intermolecular hydrogen bond with the dissociating CH$_3$OH. This additional coordination of the dissociating CH$_3$OH raises it off the surface by about $0.7$~\AA, increases its tilt $\theta$ by about $30^\circ$, and results in a reduction of the charge transferred to the O$_{\mathit{br}}$ by about $0.1e$ upon dissociation, as shown in Fig.~\ref{fgr:structures}(c) and (d).  Our aim is to see how these structural and electronic changes with coverage impact the interfacial level alignment of CH$_3$OH on TiO$_2$(110).

We begin by comparing the coverage dependence of the relevant interfacial molecular levels' spatial distribution. Specifically, in Fig.~\ref{fgr:wavefunctions} we compare the isosurfaces and height dependence of the wavefunction densities for LUMO$_{\mathrm{Wet}}$, LUMO$_{\mathrm{Wet}}$+1, HOMO$_{\mathrm{D}}$, and HOMO$_{\mathrm{ND}}$ for \nicefrac{1}{2} ML (intact and dissociated) and 1 ML (intact and \nicefrac{1}{2}-dissociated) of CH$_3$OH on TiO$_2$(110).    

\begin{table*}
\caption{\ \\DFT and $G_0W_0$ energies for the HOMO$_{\mathrm{D}}$, HOMO$_{\mathrm{ND}}$, LUMO$_{\mathrm{Wet}}$ and LUMO$_{\mathrm{Wet}}+1$ levels, highest/lowest peak energies $\varepsilon_{\mathit{peak}}^{\mathit{PDOS/Wet}}$ relative to the VBM $\varepsilon_{\mathrm{VBM}}$ for the CH$_{\mathrm{3}}$OH PDOS/Wet DOS, and differences $\Delta \varepsilon^{\mathit{UPS}/\mathit{2PP}}$ from the UPS/2PP measurements  in eV, for the \nicefrac{1}{2} ML (intact and dissociated) and 1 ML (intact and \nicefrac{1}{2}-dissociated) CH$_{\mathrm{3}}$OH on TiO$_2$(110).}\label{tbl:energies}
\resizebox{\textwidth}{!} {
\footnotesize
\begin{tabular}{cr@{ }lcccllccll}\hline
\\[-3mm]
Method & \multicolumn{2}{c}{Coverage} & Molecular 
&HOMO$_{\mathrm{D}}$ & HOMO$_{\mathrm{ND}}$
&\multicolumn{1}{c}{$\varepsilon_{\mathit{peak}}^{\mathit{PDOS}}$} 
&\multicolumn{1}{c}{$\Delta \varepsilon^{\mathit{UPS}}$} 
&LUMO$_{\mathrm{Wet}}$ & LUMO$_{\mathrm{Wet}}+1$
&\multicolumn{1}{c}{$\varepsilon_{\mathit{peak}}^{\mathit{Wet}}$} 
&\multicolumn{1}{c}{$\Delta \varepsilon^{\mathit{2PP}}$}\\
 	 & & & Layer 
& (eV) & (eV) 
& \multicolumn{1}{c}{(eV)} & \multicolumn{1}{c}{(eV)}
& (eV) & (eV) 
& \multicolumn{1}{c}{(eV)} & \multicolumn{1}{c}{(eV)}\\\hline

\multirow{4}{*}{DFT}

&\nicefrac{1}{2} & ML & Intact
& $-0.83$ & --- 
& $-0.88$ & $+0.67$ 
& $+5.51$ & ---
& $+5.73$ & $+0.15$	\\

&1 & ML& Intact	
& $-0.85$ & $-0.61$
& $-0.76^a$ & $+0.79^a$ 
& $+4.61$ & $+4.62$
& $+4.96^a$ & $-0.62^a$\\

&\nicefrac{1}{2} & ML & Dissociated 
& $-0.43$ & ---
& $-0.90$ & $+0.65$ 
& $+5.39$ & ---
& $+5.58$ & $+0.00$\\

&1 & ML& \nicefrac{1}{2}-Dissociated 
& $-0.55$ & $-0.60$
& $-0.55^a$ & $+1.00^a$ 
& $+4.95$ & $+4.96$
& $+5.27^a$ & $-0.31^a$\\\hline

\multirow{4}{*}{$G_0W_0$}
&\nicefrac{1}{2} & ML & Intact 
& $-1.28$ & ---
& $-1.31$ & $+0.24$ 
& $+6.17$ & ---
& $+6.34$ & $+0.76$ \\

&1 & ML & Intact
& $-1.88$ & $-1.32$
& $-1.29^a$ & $+0.26^a$ 
& $+5.23$ & $+5.29$
& $+5.68^a$ & $+0.10^a$\\

& \nicefrac{1}{2} & ML & Dissociated 
& $-0.65$ & ---
& $-1.05$ & $+0.50$ 
& $+6.09$ & ---
& $+6.04$ & $+0.46$\\

& 1 & ML & \nicefrac{1}{2}-Dissociated 
& $-1.03$ & $-1.05$
& $-0.97^a$ & $+0.57^a$ 
& $+5.42$ & $+5.97$
& $+5.98^a$ & $+0.40^a$\\\hline
UPS
&&&&&& $-1.55^b$
& ---   \\
2PP
&&&&&&&&&& $+5.58^c$
& --- \\
\hline
\multicolumn{6}{l}{$^a$Ref.~\citenum{OurJACS}, $^b$Ref.~\citenum{Onishi198833}, $^c$Ref.~\citenum{Onda200532}.}
\end{tabular}
}
\end{table*}

 LUMO$_{\mathrm{Wet}}$ for \nicefrac{1}{2} ML structures (Fig.~\ref{fgr:wavefunctions}(a,c)) is more hybridized with the unoccupied surface Ti 3d levels compared to the 1 ML structures (Fig.~\ref{fgr:wavefunctions}(b,d)). Further, LUMO$_{\mathrm{Wet}}$ for the dissociated \nicefrac{1}{2} ML also overlaps with the O$_{br}$H $\sigma^*$ level. For the 1 ML there is sufficient coverage for LUMO$_{\mathrm{Wet}}$ to have 2D character (\emph{cf}.\ insets Fig.~\ref{fgr:wavefunctions}(b,d)). However, \nicefrac{1}{2} ML coverage is insufficient for LUMO$_{\mathrm{Wet}}$ to be 2D (\emph{cf}.~\ insets Fig.~\ref{fgr:wavefunctions}(a,c)). Finally, LUMO$_{\mathrm{Wet}}$+1 for the \nicefrac{1}{2}-dissociated 1 ML lacks 2D character and is substantially hybridized with the surface Ti 3d levels, as we saw for the \nicefrac{1}{2} ML LUMO$_{\mathrm{Wet}}$. 

For the \nicefrac{1}{2} ML intact structure (Fig.~\ref{fgr:wavefunctions}(a)), HOMO$_{\mathrm{D}}$  is substantially hybridized with the O 2p$_{\pi}$ levels \cite{DuncanTiO2} of the surface.  For the 1 ML intact structure (Fig.~\ref{fgr:wavefunctions}(b)), HOMO$_{\mathrm{D}}$  is mostly localized on CH$_3$OH, while HOMO$_{\mathrm{ND}}$ is somewhat more hybridized with the surface. The increased localization for 1 ML is consistent with the larger separation from the surface of the dissociating CH$_3$OH shown in Fig.~\ref{fgr:structures}(c). In contrast, the reduced tilt for \nicefrac{1}{2} ML of CH$_3$OH  (Fig.~\ref{fgr:structures}(a)) results in a stronger overlap with the O 2p$_{\pi}$ levels of the three-fold-coordinated O atom (O$_{3c}$) of the surface. 

After dissociation, the HOMO$_{\mathrm{D}}$ is more hybridized with the surface for both \nicefrac{1}{2} ML and 1 ML coverages (Fig.~\ref{fgr:wavefunctions}(c,d)). The HOMO$_{\mathrm{D}}$ O 2p lone pairs are similarly oriented for both 1 ML structures. However, the HOMO$_{\mathrm{D}}$ O 2p lone pairs for the   \nicefrac{1}{2} ML intact and dissociated structures are oriented differently.  After O--H bond dissociation, the O 2p orbital, which was previously involved in the O--H bond, becomes the  least stable CH$_3$OH O 2p lone pair, i.e., HOMO$_{\mathrm{D}}$.

Table\ref{tbl:energies} lists the LUMO$_{\mathrm{Wet}}$, LUMO$_{\mathrm{Wet}}$+1, HOMO$_{\mathrm{D}}$, and HOMO$_{\mathrm{ND}}$ energy  at $\Gamma$ for \nicefrac{1}{2} ML and 1 ML of CH$_3$OH on TiO$_2$(110).

For the 1 ML intact structure, the non-dissociating CH$_3$OH pushes the dissociating molecule off the surface via steric hindrance. At the same time, their intermolecular H-bond prevents the dissociating CH$_3$OH from desorbing. This stabilizes HOMO$_{\mathrm{D}}$ relative to HOMO$_{\mathrm{ND}}$ (Table \ref{tbl:energies}). As a result, the HOMO$_{\mathrm{D}}$ energy is closer to that of the isolated CH$_3$OH (Fig.~\ref{fgr:schematic}). On the other hand, in the absence of this coordinating CH$_3$OH, the HOMO$_{\mathrm{D}}$ energy is close to the 1 ML HOMO$_{\mathrm{ND}}$ energy. 

After dissociation,  HOMO$_{\mathrm{D}}$ and HOMO$_{\mathrm{ND}}$ are shifted up to basically the same energy. This may be understood in terms of an increased interaction of the dissociated CH$_3$OH with the surface (\emph{cf}.~\ Fig.~\ref{fgr:wavefunctions}(d)). 

Overall, the unoccupied CH$_3$OH levels may be classified into two groups based on their energies (Table~\ref{tbl:energies}) and spatial distribution (Fig.~\ref{fgr:wavefunctions}).  The first group of levels is at lower energy (5.2---5.4~eV) and has a strong 2D character above the surface.  This group consists of the intact 1 ML LUMO$_{\mathrm{Wet}}$ and LUMO$_{\mathrm{Wet}}+1$ and the \nicefrac{1}{2}-dissociated 1 ML LUMO$_{\mathrm{Wet}}$ levels.   The second group of levels is at higher energy (6.0---6.2~eV), lacks 2D character, and is instead significantly hybridized with Ti 3d levels.  This group consists of the intact and dissociated \nicefrac{1}{2} ML LUMO$_{\mathrm{Wet}}$ and the \nicefrac{1}{2}-dissociated 1 ML LUMO$_{\mathrm{Wet}}+1$ levels.  

Based on the observed groupings, we find Wet electron levels with 2D character are more stable by about 0.8~eV, as compared to those which are instead hybridized with the surface.  In particular, after dissociation the 1 ML LUMO$_{\mathrm{Wet}}+1$ level loses its weight on the dissociated CH$_3$OH, and thus its 2D character.  This level is then significantly destabilized, and resembles in both energy and spatial distribution LUMO$_{\mathrm{Wet}}$ for \nicefrac{1}{2} ML coverage.  

Already at the DFT level, we find the unoccupied levels are significantly stabilized by increasing the coverage from \nicefrac{1}{2} ML to 1 ML.  Overall, the 1 ML intact structure has the most stable LUMO$_{\mathrm{Wet}}$ levels.  However, differences in the spatial distribution of the 1 ML \nicefrac{1}{2}-dissociated Wet electron levels are not reflected in their KS eigenenergies.  Capturing such effects requires a QP treatment to describe the anisotropic screening.  

Due to the differences in their spatial distribution, 2D and surface hybridized Wet electron levels of the 1 ML \nicefrac{1}{2}-dissociated structure are screened differently, resulting in different QP energy shifts.  On the one hand, the 2D LUMO$_{\mathrm{Wet}}$ QP energy approaches that of the 1 ML intact LUMO$_{\mathrm{Wet}}$ levels.  On the other hand, the surface hybridized LUMO$_{\mathrm{Wet}}+1$ QP energy approaches that of the \nicefrac{1}{2} ML LUMO$_{\mathrm{Wet}}$ levels. We have previously shown that there exists an intimate relationship between QP energy shifts and the wavefunction's spatial distribution \cite{OurJACS,MiganiLong}.

\begin{figure}[!t]
\includegraphics[width=\columnwidth]{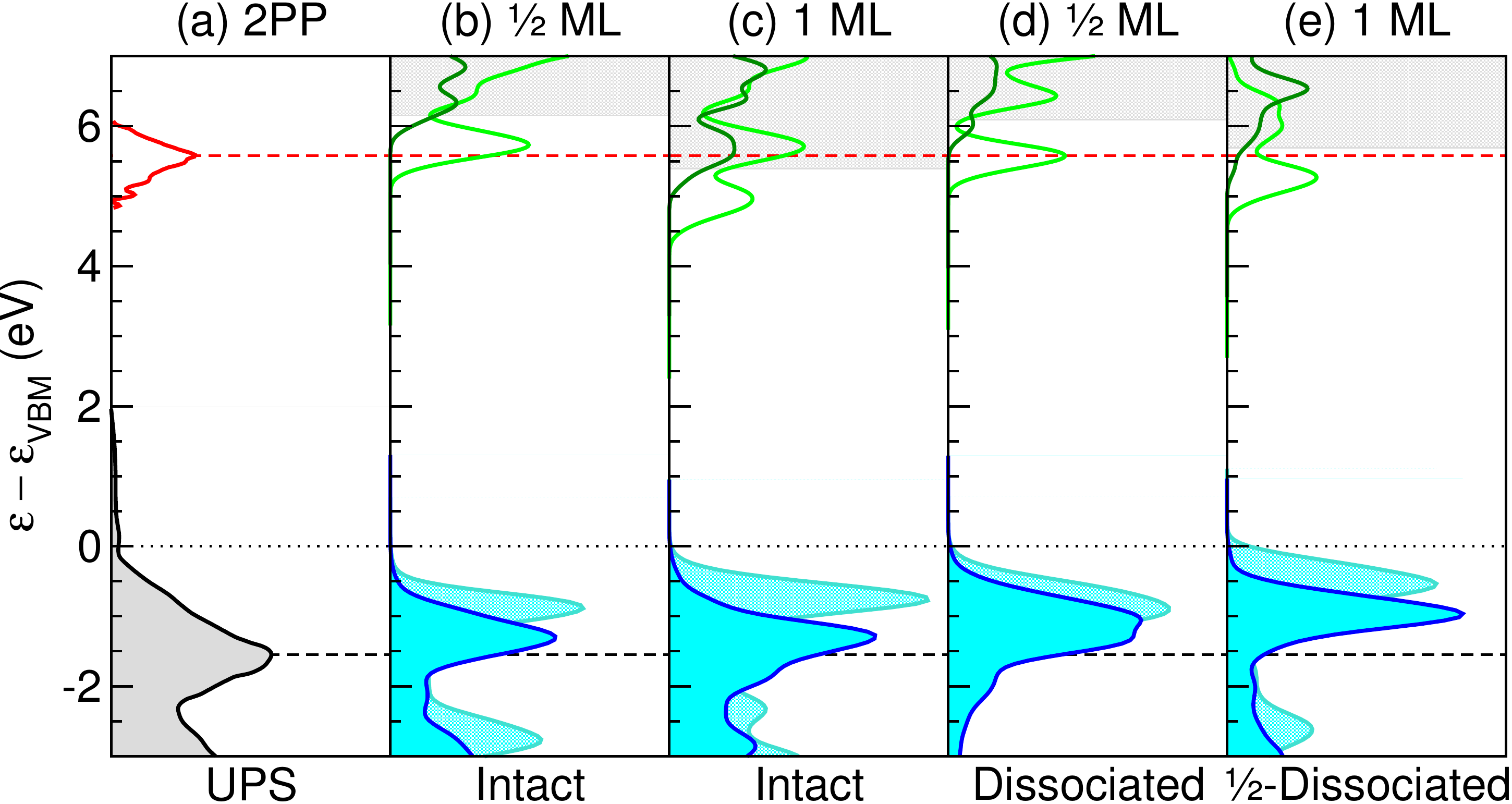}
\caption{CH$_3$OH projected (cyan/blue) and Wet (light/dark green) DOS computed with DFT/$G_0W_0$ for an (b,c) intact, (d) dissociated, and (e) \nicefrac{1}{2}-dissociated (b,d) \nicefrac{1}{2} ML and (c,e) 1 ML \cite{OurJACS} of CH$_3$OH on TiO$_2$(110) and the (a) experimental UPS \cite{Onishi198833}  (black) and 2PP spectra \cite{Onda200532} (red).  Filling denotes occupied levels. Energies are relative to the VBM, $\varepsilon_{\mathrm{VBM}}$ (black dotted line). Grey shaded regions denote states above the vacuum level $E_{vac}$.  Black/red dashed horizontal lines denote the highest/lowest energy peaks in the UPS/2PP spectra.}
\label{fgr:DOS}
\end{figure}

 As shown in Fig.~\ref{fgr:wavefunctions}, the Wet electron levels are delocalized within the molecular plane, with weight inside and above the molecular layer.  This means they may be screened based on their density averaged over the $x y$-plane.  Specifically, they may be identified as the unoccupied levels with more than half their weight between the bridging O atom of the surface and $5$~\AA\ above the top of the molecular layer. It is these levels which are included in the Wet DOS of Fig.~\ref{fgr:DOS}.

In Fig.~\ref{fgr:DOS} we compare the CH$_3$OH PDOS and Wet DOS for \nicefrac{1}{2} ML (intact and dissociated) and 1 ML (intact and \nicefrac{1}{2}-dissociated) on TiO$_2$(110) with the UPS and 2PP spectra.  We focus on the highest energy peak in the UPS, which is at $\varepsilon_{peak}^{UPS}\approx -1.55$~eV relative to the VBM, $\varepsilon_{\mathrm{VBM}}$ \cite{Onishi198833}. We find the corresponding peak in the CH$_3$OH PDOS of both 1 ML and \nicefrac{1}{2} ML intact structures is in semi-quantitative agreement with the experiment ($\Delta\varepsilon^{\mathit{UPS}}\approx0.25$~eV). The highest energy peak in the CH$_3$OH PDOS for both \nicefrac{1}{2} ML dissociated and 1 ML \nicefrac{1}{2}-dissociated structures is shifted $\sim0.54$~eV closer to the VBM compared to the UPS peak.  Thus, the $G_0W_0$ results support the assignment of the UPS highest energy peak to intact CH$_3$OH, independent of the coverage.  

Generally, the energies of the HOMO levels at $\Gamma$ coincide with the highest energy PDOS peak.  However, the CH$_3$OH PDOS for the 1 ML intact structure exhibits a pronounced shoulder at $\sim -1.9$~eV.  This feature is directly attributable to the HOMO$_{\mathrm{D}}$ at $-1.88$~eV, while the HOMO$_{\mathrm{ND}}$ contributes to the main peak at $-1.29$~eV.  

Perhaps more surprising, the HOMO$_{\mathrm{D}}$ for the dissociated \nicefrac{1}{2} ML structure is 0.5~eV above the corresponding PDOS peak.  This is observed as a broadening in the PDOS towards higher energies.  It is the counterpart of the intact HOMO$_{\mathrm{D}}$, at $-1.06$~eV, which coincides with the peak energy in the PDOS for the \nicefrac{1}{2} ML dissociated structure.  

To understand this behaviour requires some consideration of the HOMO's energy dispersion and its relation to hybridization with the surface.  The more localized a molecular level is, the less its energy dispersion.  For the dissociated \nicefrac{1}{2} ML structure, there is significant hybridization with the TiO$_2$(110) surface.  This results in broadening of the HOMO into a band as it mixes with the O 2p$_\pi$ states of the surface. 

The 2PP spectrum has an intense experimental peak at $\varepsilon_{peak}^{2PP}\approx 5.58$~eV \cite{Onda200532}. As previously reported \cite{OurJACS,MiganiLong}, we find the 2PP spectrum is best reproduced by the 1 ML CH$_3$OH intact structure ($\Delta\varepsilon^{\mathit{2PP}}\sim0.10$~eV). However, unlike the CH$_3$OH PDOS peak, we observe a qualitative difference in the Wet DOS with the coverage. In particular, the Wet DOS peak for the \nicefrac{1}{2} ML intact structure is significantly higher in energy ($\Delta\varepsilon^{\mathit{2PP}}\sim0.76$~eV).

As mentioned above, the 2D LUMO$_{\mathrm{Wet}}$ level for the 1 ML \nicefrac{1}{2}-dissociated structure resembles in energy and spatial distribution the LUMO$_{\mathrm{Wet}}$ and LUMO$_{\mathrm{Wet}}+1$ levels for the 1 ML intact structure. Consistent with its 2D character, we find the 2D LUMO$_{\mathrm{Wet}}$'s energy matches the 2PP peak. This suggests there will be LUMO$_{\mathrm{Wet}}$ levels with 2D character, which are accessible in the 2PP energy range, even for mixtures of dissociated and intact CH$_3$OH.

 All the LUMO$_{\mathrm{Wet}}$ levels contribute to the onset of the Wet DOS with the exception of LUMO$_{\mathrm{Wet}}+1$ for the 1 ML \nicefrac{1}{2}-dissociated structure. The latter contributes directly to the Wet DOS peak. 
As we saw for the HOMO levels, hybridization of the LUMO$_{\mathrm{Wet}}$ levels with the Ti 3d levels results in a broadening of these levels into bands.  As a result, we see a distinct broadening of the peaks in the Wet DOS.  This broadening is significantly more pronounced for the \nicefrac{1}{2} ML LUMO$_{\mathrm{Wet}}$ levels, due to their stronger hybridization with the surface.

\begin{figure}[!t]
\includegraphics[width=\columnwidth]{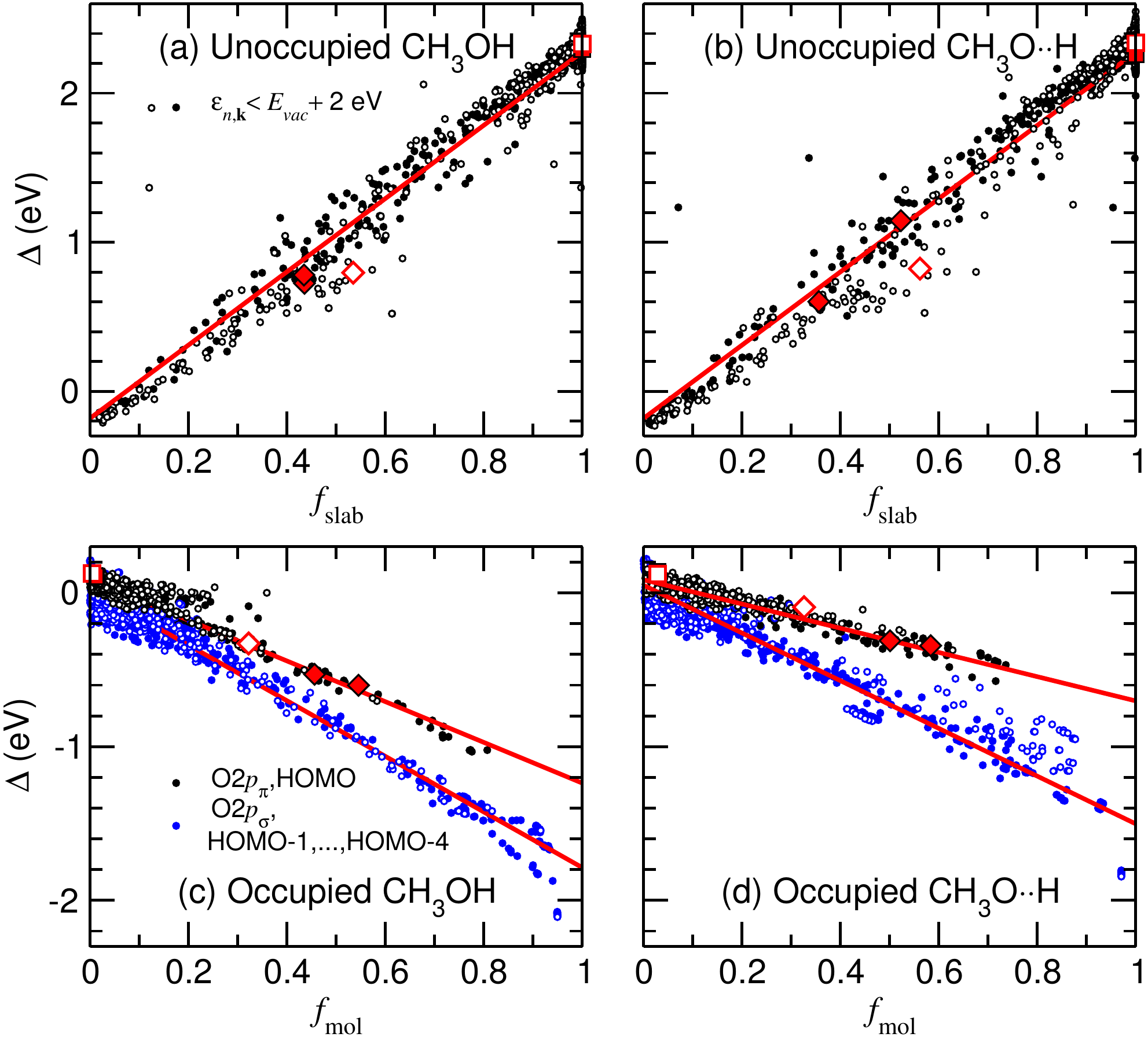}
\caption{$G_0W_0$ QP energy correction $\Delta$ in eV versus fraction of the wavefunction's density (a,b) in the slab $f_{\mathrm{slab}}$ for the unoccupied levels and (c,d) in the molecular layer $f_{\mathrm{mol}}$ for the occupied levels with \nicefrac{1}{2}~ML (open symbols) and $1$~ML \cite{OurJACS} (filled symbols) of
(a,c) intact and (b,d) either dissociated or \nicefrac{1}{2}-dissociated CH$_3$OH on TiO$_2$(110).  Diamonds denote the (a,b) LUMO$_{\mathrm{Wet}}$, LUMO$_{\mathrm{Wet}}+1$, (c,d) HOMO$_{\mathrm{D}}$, HOMO$_{\mathrm{ND}}$ at $\Gamma$ depicted in Fig.~\ref{fgr:wavefunctions}.  Squares denote the (a,b) CBM and (c,d) VBM levels. Red solid lines are linear fits.  
}
\label{fgr:correlations}
\end{figure}

To understand the nature of the observed differences in the QP $G_0W_0$ corrections to the KS eigenenergies, we consider their relation to the spatial distribution of the KS wavefunction. In Fig.~\ref{fgr:correlations} we plot the calculated QP $G_0W_0$ energy corrections $\Delta$ versus the fraction of the wavefunction within the surface and molecular layer $f_{\mathrm{slab}}$ (grey and brown regions in Fig.~\ref{fgr:wavefunctions})  for the unoccupied levels, and within the molecular layer $f_{\mathrm{mol}}$ (brown regions in Fig.~\ref{fgr:wavefunctions}) for the occupied layers.  We find the same linear correlations, independent of the CH$_3$OH coverage, as reported in Refs.~\citenum{OurJACS,MiganiLong}.  

These results suggest that for a molecular level to be available in the 2PP energy regime, it should have little weight in the bulk, i.e., a weak hybridization with the Ti 3d levels.  Otherwise, the level would be shifted above the vacuum level.  Instead, any accessible level should have significant weight in the vacuum.  In effect, any molecular level that meets these criteria could be reasonably classified as a Wet electron level.

Altogether, this suggests these linear correlations may be effectively used to estimate the QP $G_0W_0$ eigenenergies based on the KS wavefunction's spatial distribution.  This would pave the way for future studies of more complicated CH$_3$OH overlayers, and lower coverages, for which $G_0W_0$ calculations are computationally unfeasible \cite{OurJACS}.

\section{Conclusions}

We have performed $G_0W_0$ calculations for the \nicefrac{1}{2} ML (intact and dissociated) and 1 ML (intact and \nicefrac{1}{2}-dissociated) CH$_3$OH on TiO$_2$(110) structures.  These results provide an accurate description of the HOMO/LUMO$_{\mathrm{Wet}}$ level alignment on TiO$_2$(110) as a function of the coverage.  It is this interfacial level alignment which controls the photocatalytic activity of the system \cite{HendersonSurfSciRep,YatesChemRev,MiganiLong}.

By comparing our $G_0W_0$ results to those from standard DFT, we find that the interfacial screening impacts the level alignment qualitatively.  This is particularly evident for the LUMO$_{\mathrm{Wet}}$ levels of the 1 ML \nicefrac{1}{2}-dissociated structure, which have different spatial distributions.  Although they have have the same KS energy, their QP energies are $\sim 0.55$ eV apart. This is because differences in screening of the wavefunctions are not captured by standard DFT.

We have shown that it is the spatial distribution of the KS wavefunction which determines the QP energy shifts.  This is because the wavefunction is screened differently in the three interfacial regions: bulk, molecular, or vacuum.  This makes a QP treatment of the anisotropic screening necessary for an interfacial system. In summary, the wavefunction's spatial distribution controls the QP level alignment.

 We find the HOMO wavefunction becomes more hybridized with the substrate as the coverage decreases.  As a result, the HOMO energy spans a wide range (1.2~eV) as a function of coverage and dissociation.
 The HOMO for \nicefrac{1}{2} ML dissociated CH$_3$OH on TiO$_2$ is the least stable amongst the structures we have considered, at $-0.65$~eV below the VBM.  

 This energy suggests that there may exists a nearby nuclear configuration at which a realignment of the energy levels may occur. Further, the strong hybridization of the CH$_3$O HOMO with the surface O 2p levels suggests a facile hole transfer between these levels.
 These two points together imply that a hole generated in the TiO$_2$ surface via photoexcitation may be transferred to CH$_3$O.  This suggests isolated CH$_3$O should be a better hole trap than CH$_3$OH.

 Overall, there are two types of LUMO$_{\mathrm{Wet}}$ levels: those with 2D character and weight above the surface, and those with a significant hybridization with the surface.
 At higher coverages, we find a preponderance of 2D LUMO$_{\mathrm{Wet}}$ levels, while at lower coverages, the LUMO$_{\mathrm{Wet}}$ levels are significantly hybridized with the surface.

 The experimentally observed 2PP resonance appears to arise from excitations to 2D LUMO$_{\mathrm{Wet}}$ levels.
 This is contrary to the previous assignment ``to the perturbed Ti 3d orbitals caused by the strong interaction of Ti ions with the CH$_3$O species'' \cite{MethanolSplitting2010,Methanol2PPYang}.

However, one should bear in mind that $G_0W_0$ tends to over-correct the Ti 3d levels, i.e., the TiO$_2$ conduction band minimum (CBM) \cite{MiganiLong}.  This could impact the calculated Wet DOS peak positions for LUMO$_{\mathrm{Wet}}$ levels as they are significantly hybridized with the Ti 3d levels.  

A correct description of the Ti 3d levels, and hence the CH$_3$OH--TiO$_2$(110) interface, may require a self-consistent $GW$ treatment, including the ionic polarization within the dielectric response function \cite{Marques}. 
Finally, one should solve the Bethe-Salpeter Equation (BSE) for a reduced TiO$_2$(110) surface to demonstrate whether excitations from reduced Ti$^{+3}$ 3d levels to the Wet electron levels have a high oscillator strength \cite{HybertsenTiO2GW,AmilcareTiO2GW,JuanmaRenormalization2,VitoBenzeneBSE}.

\section*{Acknowledgements}

We acknowledge fruitful discussions with Amilcare Iacomino, Jin Zhao, Hrvoje Petek, and Angel Rubio; funding from the Spanish Grants (FIS2012-37549-C05-02, RYC-2011-09582, JCI-2010-08156); and computational time from BSC Red Espanola de Supercomputacion.
\section*{References}

\bibliography{Bibliography}

\end{document}